# Carnot's theorem and Szilárd engine

Liangsuo Shu[1,2], Xiaokang Liu[1], Suyi Huang[1], Shiping Jin[1,2,3]

[1]School of energy and power engineering, Huazhong University of Science & Technology. Wuhan, China.
[2]Innovation Institute, Huazhong University of Science & Technology. Wuhan, China.
[3]China-Europe Institute for Clean and Renewable Energy, Huazhong University of Science & Technology. Wuhan, China.

**Abstract**

In this work, the relationship between Carnot engine and Szilárd engine was discussed. By defining the available information about the temperature difference between two heat reservoirs, the Carnot engine was found to have a same physical essence with Szilárd engine: lossless conversion of available information. Thus, a generalized Carnot's theorem for wider scope of application can be described as "all the available information is 100% coded into work".

**Introduction**

As one of the most fundamental physics laws, the second law of thermodynamics can trace its origin back to Carnot's theorem obtained by Sadi Carnot in 1824, which limits the maximum efficiency for any possible engine. However, James Clerk Maxwell created a thought experiment (Maxwell's demon) in which the second law might be violated in 1867. In 1929, Leó Szilárd invented an engine which can transform heat from an isothermal environment into work at the cost of the information consumption of a Maxwell's demon[1]. After the efforts of many researchers[2–5], Maxwell's demon was proved to be consistent with the second law. In recent year, benefit from advances in technology, various kinds of Maxwell's demons have been realized in laboratories with methods of both physics[6,7] and chemistry[8–10]. In biology, the special channel structure of aquaporin-1 was also found to enable it to work as a suspect Maxwell's demon by recognizing and consuming the information about the size difference between different solute molecules[11]. More theoretical schemes of heat engine based on Maxwell's demon have been put forward[5,12–17]. However, there is still much controversy on whether these kinds of heat engines (especially the quantum ones) respect Carnot's theorem[15,16,18] or not[19–22].

In a previous work, we have generalized the principle of entropy increase for isolated system to the

liangsuo_shu@hust.edu.cn

principle of available information decrease for any system. The available information of an observer about a system is the quantificational measurement of the causes of any observable effects. In this work, we analyze the available information in the Carnot engine and Szilárd engine and find that they two have a same physical essence: lossless conversion of available information. Carnot's theorem was generalized as "the available information is 100% coded into work".

**Available information in Carnot's cycle and Szilárd engine**

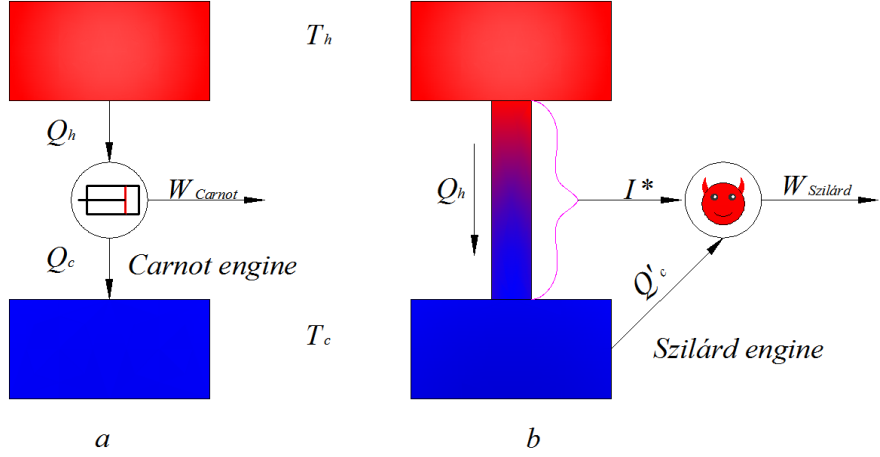

**Fig.1** Forward cycle of Carnot engine(*a*) and Szilárd engine(*b*). For Szilárd engine, a Maxwell's demon extracts this $I^*$ and codes it in the heat pumped from the environment ($Q'_c$) to output work($W_{Szilárd}$); For Carnot engine, the isothermal endotherm is a process of writing available information of $I^*$, which is coded in output work ($W_{Carnot}$) subsequently and rejects the waste heat of $Q_c$ without available information.

Supposing there are two heat reservoirs, a hot one and a cold one, with temperature of $T_h$ and $T_c$ respectively, for a Carnot engine (**Fig**.1.*a*) driven by an amount of heat ($Q_h$) from hot reservoir, the maximum work $W_{Carnot}$ will be,

$$W_{Carnot} = Q_h \eta_{Carnot} = Q_h(1 - \frac{T_c}{T_h}) \qquad (1)$$

where $\eta_{Carnot}$ is the Carnot efficiency, the maximum efficiency for any possible engine according to the second law of thermodynamics. The heat got by cold reservoir will be,

$$Q_c = Q_h - W_{Carnot} = Q_h \frac{T_c}{T_h} \qquad (2)$$

The available information about the temperature difference carried by the amount of heat $Q_h$ at $T_h$
liangsuo_shu@hust.edu.cn

can be described as below[1],

$$I^*_{Q_h} = \frac{Q_h}{k}(\frac{1}{T_c} - \frac{1}{T_h}) \quad (3)$$

In a heat transfer without any work output, $I^*_{Qh}$ will be consumed and results in an entropy increase of $kI^*_{Qh}$. If $I^*_{Qh}$ serves as the information source of the Maxwell's demon in a Szilárd engine (**Fig**.1*b*) in cold reservoir, the heat pumped from the cold reservoir and transformed into work will be,

$$W_{Szilard} = kT_c I^*_{Q_h} = Q_h(1 - \frac{T_c}{T_h}) \quad (4)$$

Comparing Eq. (1) and (4), we can find that the Szilárd engine obeys Carnot's theorem. In a cycle, the available information of heat reservoir decreases,

$$\Delta I^*_h = -I^*_{Q_h} = -\frac{Q_h}{k}(\frac{1}{T_c} - \frac{1}{T_h}) < 0 \quad (5)$$

the available information of cold reservoir remains unchanged,

$$\Delta I^*_c = \frac{Q_c}{k}(\frac{1}{T_c} - \frac{1}{T_c}) = 0 \quad (6)$$

All the available information about the temperature difference is taken away by the work, leaving no available information in $Q_c$. Comparing **Eq**. (4), (5) and (6), Carnot's theorem can be described in the language of available information as,

"*the available information about the temperature difference between two heat* reservoirs *carried by the heat is 100% coded into work*".

For Szilárd engine here, in order to make the Maxwell's demon able to get all of $I^*$, the heat transfer of $Q_h$ from $T_h$ to $T_c$ must be a reversible process experiencing an infinite number of intermediate heat sources to avoid additional consumption of available information during heat transfer. The net heat obtained by cold reservoirs will be,

---

$$Q_c = Q_h - Q_c^{'} = Q_h - W_{Szilard} = Q_h \frac{T_c}{T_h} \tag{7}$$

Therefore, after a complete cycle of a Carnot engine or a Szilárd engine, the states of the two heat reservoirs are the same. In fact, if both of them are placed in black boxes, an outside observer will find no method to distinguish them.

Heat engine connecting heat reservoirs in non-equilibrium states had been discussed by Hasegawa et al. [23] and Takara et al.[24], and were found to be able to output more work than those connecting equilibrium states. In fact, beside the available information about temperature difference describing the imbalance between system and its environment, the internal imbalance of a system is also one kind of available information. For Szilárd engine, this internal available information of hot heat reservoirs can also act as part of information source while some residual available information must be retained as the origin of the internal available information of cold reservoirs. For a system composed of nearly-ideal gas of molecules, its internal available information can be described by Boltzmann's $H$ function[2]. Therefore, the possible maximum available information coded in work in the cycle will be,

$$I^{*}_{non-equilibrium} = I^{*} + H_h - H_c \tag{8}$$

where $H_h$ and $H_c$ are the $H$ of the hot and the cold reservoirs respectively. If $H_c=0$ (final state is an equilibrium state), the work will take its maximum value($W_{max(non-eq)}$), which can be greater than the value of a classic Carnot engine (**Eq**.(1) and **Eq**.(4)).

$$W_{max(non-eq)} = kT_c(I^{*} + H_h) \tag{9}$$

Beside temperature difference, other kinds of internal and external imbalances are also available information. Therefore, Carnot's theorem can be generalized as,

*"All the available information is 100% coded into work"*.

Classic Carnot engine cannot utilize $H$ of a non-equilibrium state. In the case which local equilibrium assumption can be met, one possible approach to utilize it is to split the hot heat source into a number of internal balanced subsystems between which many small Carnot engines or molecular information ratchets[8–10] can run. They cannot violate the generalized Carnot's theorem above.

**Reversed cycle**

If $Q_c$ is moved from a cold reservoir to a hot reservoir, the hot reservoir will get both $Q_c$ and the available information about the temperature difference between the cold reservoir and itself carried by $Q_c$, which must be replenished from outside because $Q_c$ from the cold reservoir has no available information (**Eq.** (6)).

$$\Delta I_h^* = I_{Q_c}^* = \frac{Q_c}{k}(\frac{1}{T_c} - \frac{1}{T_h}) \tag{10}$$

For the reversed cycle of both the Carnot engine (**Fig**.2.*a*) and Szilárd engine (**Fig**.2.*b*), this replenishment can be achieved through external work ($W'$). For Carnot engine, the work is,

$$W'_{Carnot} = Q_c(\frac{T_h}{T_c} - 1) \tag{11}$$

For Szilárd engine, the work is

$$W'_{Szilard} = kT\Delta I_h^* \tag{12}$$

where $T$ is the operating temperature of information writing process. when $T$ is equal to $T_h$, $W'_{Szilárd}$ will take its minimum value (if $T$ is lower than $T_h$, the information writing process cannot be completed).

$$W'_{Szilard-\min} = kT_h\Delta I_h^* = Q_c(\frac{T_h}{T_c} - 1) \tag{13}$$

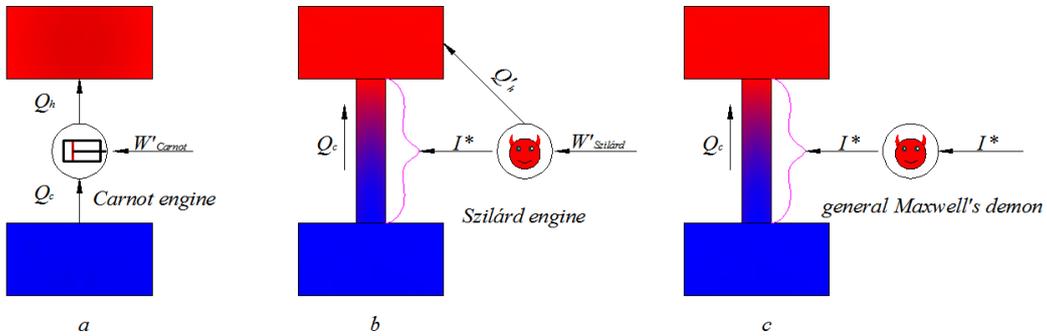

**Fig.2** Reversed cycle of Carnot engine(*a*), Szilárd engine(*b*) and general Maxwell's demon.

In the above two cases, the total heat got by the hot reservoir will be $Q_h= Q_cT_h/T_c$. Therefore, in a reversed cycle, an outside observer still cannot distinguish Szilárd engine from Carnot engine. For
liangsuo_shu@hust.edu.cn

general Maxwell's demon (**Fig**.2.*c*) beside Szilárd engine, the replenishment can also be achieved by consuming other available information such as the information stored in DNA or a computer's hard drive and the information about the difference between ATP and ADP.

**Conclusion and discussion**

From the discussion above, it can be found that the physical essence of Carnot engine and Szilárd engine is the same: 100% transmission of available information. Compared with equal amount of heat at the same temperature of *T*, work carries additional available information,

$$W(Q, I^*) = Q + I^* \tag{14}$$

$$I^* = \frac{W}{kT} \tag{15}$$

where *T* is the information operating temperature. For a forward cycle of both Carnot engine and Szilárd engine, $I^*$, carried by $Q_h$ about the temperature difference between hot and cold reservoir, is encoded in work; In a reversed cycle, $I^*$ from work act as the available information source to cover the shortfall of available information. From **Eq.** (4), (10) and (15), it can be found that the amount of information carried by work changes with the operating temperature. This means that the stored information can also be used as a "working medium" of a cycle. **Fig**.1*a* has shown the available information process of Carnot engine from the view of heat reservoir regarding the engine as a black box and taking no account of kinetic details of the cycle. In the supplementary materials, this problem was analyzed from the view of Carnot engine and the generalized Carnot's theorem is found to still be followed.

In a complete cycle, a heat engine based on Maxwell's demon, whether a classic or quantum one, can only use existing available information (already stored in demon's memory or extracted from outside information source) but unable to create any new available information. Therefore, no one can violate the generalized Carnot's theorem (at least statistically).

**Reference**
[1]     L. Szilard, On the decrease of entropy in a thermodynamic system by the intervention of intelligent beings, Behav. Sci. 9 (1964) 301–310. doi:10.1002/bs.3830090402.
[2]     R. Landauer, Irreversibility and Heat Generation in the Computing Process, IBM J. Res. Dev. 5 (1961) 183–191. doi:10.1147/rd.53.0183.

liangsuo_shu@hust.edu.cn

[3]     C.H. Bennett, The thermodynamics of computation-a review, Int. J. Theor. Phys. 21 (1982) 905–940. doi:10.1007/BF02084158.

[4]     T. Sagawa, M. Ueda, Minimal energy cost for thermodynamic information processing: Measurement and information erasure, Phys. Rev. Lett. 102 (2009) 1–4. doi:10.1103/PhysRevLett.102.250602.

[5]     K. Maruyama, F. Nori, V. Vedral, Colloquium: The physics of Maxwell's demon and information, Rev. Mod. Phys. 81 (2009) 1–23. doi:10.1103/RevModPhys.81.1.

[6]     S. Toyabe, T. Sagawa, M. Ueda, E. Muneyuki, M. Sano, Experimental demonstration of information-to-energy conversion and validation of the generalized Jarzynski equality, Nat. Phys. 6 (2010) 988–992. doi:10.1038/nphys1821.

[7]     A. Bérut, A. Arakelyan, A. Petrosyan, S. Ciliberto, R. Dillenschneider, E. Lutz, Experimental verification of Landauer's principle linking information and thermodynamics, Nature. 483 (2012) 187–189. doi:10.1038/nature10872.

[8]     A. Carlone, S.M. Goldup, N. Lebrasseur, D.A. Leigh, A. Wilson, A Three-compartment Chemically-driven Molecular Information Ratchet, J. Am. Chem. Soc. 134 (2012) 8321.

[9]     V. Serreli, C.-F. Lee, E.R. Kay, D. a Leigh, A molecular information ratchet., Nature. 445 (2007) 523–527. doi:10.1038/nature05452.

[10]    J.P.S. Peterson, R.S. Sarthour, A.M. Souza, I.S. Oliveira, J. Goold, K. Modi, et al., Experimental demonstration of information to energy conversion in a quantum system at the Landauer limit, Proc. R. Soc. A Math. Phys. Eng. Sci. 472 (2016) 20150813. doi:10.1098/rspa.2015.0813.

[11]    L. Shu, Y. Li, Xiaokang, L.X. Qian, S. Huang, S. Jin, et al., Aquaporin-1 is a Maxwell's Demon in the Body, (2015) 1–25. http://arxiv.org/abs/1511.07177.

[12]    P. Strasberg, G. Schaller, T. Brandes, M. Esposito, Thermodynamics of a Physical Model Implementing a Maxwell Demon, Phys. Rev. Lett. 110 (2013) 40601. doi:10.1103/PhysRevLett.110.040601.

[13]    D. Mandal, C. Jarzynski, Work and information processing in a solvable model of Maxwell's demon, Proc. Natl. Acad. Sci. 109 (2012) 11641–11645. doi:10.1073/pnas.1204263109.

[14]    A.M. Jayannavar, Simple model for Maxwell's-demon-type information engine, Phys. Rev. E. 53 (1996) 2957–2959. doi:10.1103/PhysRevE.53.2957.

[15]    M.O. Scully, M.S. Zubairy, G.S. Agarwal, H. Walther, Extracting Work from a Single Quantum Coherence, Science (80-. ). 299 (2003) 862–865. doi:10.1126/science.1078955.

[16]    M.O. Scully, Extracting work from a single thermal bath via quantum negentropy., Phys. Rev. Lett. 87 (2001) 220601. doi:10.1103/PhysRevLett.87.220601.

[17]    H.J. Jeon, S.W. Kim, Optimal work of the quantum Szilard engine under isothermal processes with inevitable irreversibility, New J. Phys. 18 (2016) 0. doi:10.1088/1367-2630/18/4/043002.

[18]    T.D. Kieu, The second law, Maxwell's demon, and work derivable from quantum heat engines, Phys. Rev. Lett. 93 (2004) 1–4. doi:10.1103/PhysRevLett.93.140403.

[19]    T. Sagawa, M. Ueda, Second Law of Thermodynamics with Discrete Quantum Feedback Control, Phys. Rev. Lett. 100 (2008) 80403. doi:10.1103/PhysRevLett.100.080403.

[20]    S. Hormoz, Quantum collapse and the second law of thermodynamics, Phys. Rev. E - Stat. Nonlinear, Soft Matter Phys. 87 (2013) 1–9. doi:10.1103/PhysRevE.87.022129.

[21]    M. Esposito, G. Schaller, Stochastic thermodynamics for 'Maxwell demon' feedbacks, EPL (Europhysics Lett. 99 (2012) 30003. doi:10.1209/0295-5075/99/30003.

liangsuo_shu@hust.edu.cn

[22] E. Boukobza, H. Ritsch, Breaking the Carnot limit without violating the second law: A thermodynamic analysis of off-resonant quantum light generation, Phys. Rev. A - At. Mol. Opt. Phys. 87 (2013) 1–6. doi:10.1103/PhysRevA.87.063845.

[23] H.H. Hasegawa, J. Ishikawa, K. Takara, D.J. Driebe, Generalization of the second law for a nonequilibrium initial state, Phys. Lett. Sect. A Gen. At. Solid State Phys. 374 (2010) 1001–1004. doi:10.1016/j.physleta.2009.12.042.

[24] K. Takara, H.H. Hasegawa, D.J. Driebe, Generalization of the second law for a transition between nonequilibrium states, Phys. Lett. Sect. A Gen. At. Solid State Phys. 375 (2010) 88–92. doi:10.1016/j.physleta.2010.11.002.

**Supplementary materials**

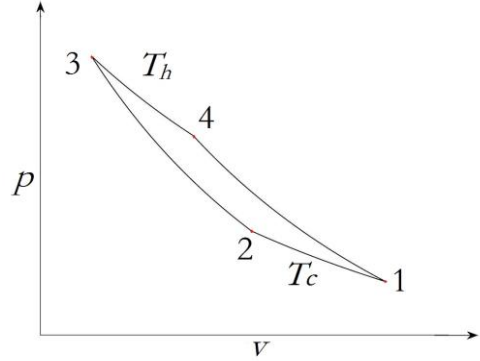

**Fig.s1** Carnot cycle

A complete Carnot cycle consists of following four steps: isothermal heat rejection (1→2), isentropic compression (2→3), isothermal heat absorption (3→4), and isentropic expansion (4→1). From the view of Carnot engine, the work from the environment will bring available information into it, while the output work to environment will take away available information.

$$dI_W^* = \frac{dW}{kT} = \frac{pdv}{kT} \quad \text{(s.1)}$$

where $I_W^*$ is the available information carried by work. Assuming the working medium is ideal single-atom gas, putting ideal gas equation of state (*pv=nRT*) into the above equation and integrating it along the process of isothermal heat rejection (1→2) and isentropic compression (2→3), we can get,

$$I_{W_{in}}^* = nR(\ln\frac{v_2}{v_1} + \ln\frac{v_3}{v_2}) = nR\ln\frac{v_3}{v_1} \quad \text{(s.2)}$$

During the isothermal heat absorption, $I_{W_{in}}^*$ is heated, which will be coded in the output work. integrating **Eq**. (s.1) along the process of isothermal heat absorption (3→4) and isentropic expansion (4→1), we will get

liangsuo_shu@hust.edu.cn

$$I^*_{W_{out}} = nR(\ln\frac{v_4}{v_3} + \ln\frac{v_1}{v_4}) = nR\ln\frac{v_1}{v_3} \qquad (s.3)$$

Comparing **Eq**. (s.3) and **Eq**. (s.2), it can be found that,

$$I^*_{W_{out}} = -I^*_{W_{in}} \qquad (s.4)$$

**Eq**. (s.4) shows that from the view of Carnot engine, the input and output available information are equal during a complete cycle (the negative sign indicates the direction is different). Therefore, the generalized Carnot's theorem still holds.

For an Otto cycle, the heat absorption and rejection of which are achieved at a constant-volume, the input and output process and "heat treatment process" of available information are completed separately. From the view of an engine, a thermodynamic cycle consists of four links of available information: inputting, heating, outputting and erasing residue to reset. The first three links can be mixed, while the last link is not necessary to ideal cycles such as Carnot cycle.

liangsuo_shu@hust.edu.cn